\documentclass[11pt,a4paper]{article}
\usepackage{amssymb}
\usepackage{graphicx}
\usepackage{amsmath}
\usepackage{makeidx}
\usepackage{indentfirst}
\usepackage[T1]{fontenc}
\usepackage[utf8]{inputenc}
\usepackage{authblk}

\newcounter{resultnum}[section]
\newcounter{conclusionnum}[section]
\newcounter{conditionnum}[section]
\newcounter{conjecturenum}[section]
\newcounter{examplenum}[section]
\newcounter{exercisenum}[section]
\newcounter{lemmanum}[section]
\newcounter{notationnum}[section]
\newcounter{theoremnum}[section]
\newcounter{definitionnum}[section]
\newcounter{corollarynum}[section]
\newcounter{remarknum}[section]
\newcounter{propositionnum}[section]
\newcounter{acknowledgementnum}[section]
\newcounter{algorithmnum}[section]
\newcounter{axiomnum}[section]
\newcounter{casenum}[section]
\newcounter{claimnum}[section]
\newcounter{summarynum}[section]
\newcounter{problemnum}[section]
\begin{document}

\title{Black Ring and Kerr Ellipsoid -- Solitonic Configurations\\
in Modified Finsler Gravity }
\date{Version accepted by IJGMMP on May 25, 2015;\ DOI: 10.1142/S0219887815501029}
\author{
\small Subhash Rajpoot\\
\vspace{-.15in}
\small\it California State University at Long Beach, \\
\small\it Long Beach, California, USA  \\
\small\it email: Subhash.Rajpoot@csulb.edu  \\
${}$ \\
Sergiu I. Vacaru\\
\small\it University "Al. I. Cuza" Ia\c si, Rector's Department  \\
\small\it 14 Alexadnru Lapu\c sneanu street, Corpus R, UAIC, office 323, Ia\c si, Romania 700057  \\
\small and \\
\small \it Max-Planck-Institute for Physics, Foehringer Ring 6,  Muenchen,  Germany D-80805\footnote{DAAD fellowship  affiliation 1} \& \\
\small \it  Inst.  Theor. Phys., Lebiniz Univ. Hannover, Appelstrasse 2, Hannover, Germany, D-30167\footnote{DAAD fellowship  affiliation 2} \\
\small\it email: sergiu.vacaru@uaic.ro ; Sergiu.Vacaru@gmail.com}

\maketitle

\begin{abstract}
We study an effective Einstein--Finsler theory on tangent Lorentz bundle constructed as a "minimal" extension of general relativity.  Black ring and  Kerr like ellipsoid exact solutions and soliton configurations are presented. In this endeavor the relevant metric depends not only on four dimensional spacetime coordinates and also on velocity type variables that can be interpreted as additional coordinates in the space of "extra dimensions".

\vskip5pt


\vskip5pt

\end{abstract}


\renewcommand\Authands{ and }

\section{Introduction}

Black rings do not exist in the classical general relativity (GR). This provides a strong motivation for conducting studies in  dimensions greater than four. The other motivation, equally strong, is string/M-theory. Such a theory is a promising candidate for providing a unified theory of particle interactions and aims to incorporate a consistent theory of quantum gravity.

Originally, a rotating black ring was demonstrated to exist as an exact solution of the vacuum Einstein equations in five dimensional gravity for a black hole with an event horizon  topology of $ S^{1}\times S^{2},$ which is stationary and asymptotically flat \cite{1,2,emp}. It is an example of a metric with non--spherical horizon topology and a counterexample  to black hole uniqueness theorem \cite{emp1}. Since then, many new black ring solutions have been discovered, with the inverse scattering method (ISM) playing a central role in this process. The first black ring solution constructed with the ISM was that of Pomeransky and Sen’kov \cite{15}, generalizing the Emparan-Reall
black ring \cite{emp}  to include a second angular momentum. Following \cite{15}, other (concentric) doubly spinning,
asymptotically flat black rings in five dimensions followed. These include black saturns \cite{16}, di-rings \cite{17,18} and bicycling
 black rings \cite{19}. More recent inverse-scattering constructions include the explicit derivation of the unbalanced Pomeransky-Sen’kov black ring \cite{20,21}, and rings on topologically non-trivial backgrounds  \cite{22,23}. For completeness, we also state that black rings and other extended objects  relevant to black ring solutions have also been constructed in supersymmetric theories \cite{A,B,BB,C,D,E,F,G}.

Briefly, in general terms, the line element in Finsler geometry is defined as
\begin{equation}\label{eqn:finslerlength}
s[x] = \int d\tau\,F(x(\tau),\dot{x}(\tau)).
\end{equation}
The Finsler fundamental generating function F is a function on the tangent bundle \(TM\) of a manifold \(M\). Given coordinates \((x^a)\) on \(M\), one  defines coordinate basis  of \(TM\) as \(\partial/\partial y^a\). The coordinates \((x^a,y^a)\) on \(TM\) are called total bundle coordinates.
Geometry of spacetimes is generalized from the widely used notion of Finsler geometries with Euclidean signature to the Lorentzian signature \cite{vaxiom,stavrv1}.
The  geodesic equation can be written in arc length parametrization as
\begin{equation}\label{eqn:geodesic}
\ddot{x}^a + N^a{}_b(x,\dot{x})\dot{x}^b = 0\,,
\end{equation}
where \(N^a{}_b\) denotes the coefficients of the the {Cartan non-linear connection}. In Finsler spacetime,  \(N^a{}_b\) is not the usual Levi-Civita connection. The Cartan non-linear connection induces a unique split of the tangent bundle \(TTM\) with basis $ (\delta_a, \frac{\partial}{\partial y^a} )$ into horizontal and vertical parts, \(TTM = HTM \oplus VTM\). The horizontal tangent bundle \(HTM\) is spanned by the vector fields
\begin{equation}
\{\delta_a = \frac{\partial}{\partial x^a} - N^b{}_a\frac{\partial}{\partial y^b}\}\,,
\end{equation}
while the vertical tangent bundle \(VTM\) is spanned by $\frac{\partial}{\partial y^a}$. The dual basis of \(T^*TM\) is given by
\begin{equation}
\{dx^a, \delta y^a = dy^a + N^a{}_bdx^b\}\,.
\end{equation}
The tangent bundle \(TM\) of a Finsler spacetime is equipped with a metric $\mathbf{g}$ called the \emph{Sasaki metric}, which can most conveniently be expressed  as
\begin{equation}\label{eqn:sasakimetric}
\mathbf{g} = g^F_{ab}dx^a \otimes dx^b + g^F_{ab}\delta y^a \otimes \delta y^b\,.
\end{equation}
 An added bonus is the fact that the horizontal and vertical tangent spaces are mutually orthogonal with respect to the Sasaki metric. We adapt this formalism to our work in the following.

We study black ring solutions  on the tangent bundle $TM_{1}^{3}$ to a 4D Lorentz manifold $M_{1}^{3}$ with signature $(+,+,-,+)$ using the so--called anholonomic frame deformation method (AFDM). This geometric method admits  generic off--diagonal exact solutions in GR and modified gravity theories (MGTs) \cite {vkerrhd,odintsov,capozzello}. Such  extensions of GR are in effect   Einstein--Finsler gravity (EFG) theories.
Here we use the term Einstein--Finsler gravity  in a more general
context when  GR is naturally extended from $M_{1}^{3}$ to $TM_{1}^{3}$
using the same axiomatic scheme as in the Einstein gravity theory but for
certain metric compatible and N--adapted linear connections defined
canonically by the same metric tensor in the total space and arbitrary modified dispersion relations (MDRs). The velocity type coordinates are effectively treated as extra dimension coordinates and can be
geometrized on tangent bundles endowed with nonholonomic distributions
determined by MDRs.

The paper is organized as follows. In section \ref{s2}, we show how MDR
determine canonical lifts of the geometric/ physical objects and fundamental
field equations from the base Lorentzian manifolds to total spaces of
respective (co) tangent bundles. We also apply the AFDM for constructing
generic off--diagonal stationary solutions and noholonomic deformations of
conventional 4D and 8D locally anisotropic spacetimes. In section \ref%
{s3}, two classes of exact solutions for analogous black ring and Kerr black
holes on tangent Lorentz bundles are constructed using the AFDM. We study
off--diagonal deformations and nonlinear superpositions of the black ring
and black hole metrics in section \ref{s4}.  Finally, section \ref{s5} is devoted to
concluding remarks.

\section{Gravity on Tangent Lorentz Bundles}
\label{s2}  General relativity  is conventionally geometrized on a Lorentz
manifold $M_{1}^{3}$, i.e on a 4D pseudo--Riemannian spacetime, using the
principle of equivalence together with the postulates of special relativity
 which  hold true at any point in $u\in M_{1}^{3}$. The
metric  $\mathbf{g}(u)$ and other fundamental fields, geometric or otherwise,  depend explicitly  on spacetime coordinates $u=\{(x^{i},y^{a})=(x^{%
\widehat{i}},y^{3}=t)\}.
$\footnote{%
Indices are  labelled in the form $\widehat{i},\widehat{j},\widehat{k},...1,2,4$
and $i,j,k,...=1,2;a,b,c...=3,4$ corresponding to signature $(+,+,-,+),$ and  $%
y^{3}$ is fixed as the timelike coordinate.
Coordinates are also split into either $3+1$ or $2+2$ splitting as needed.}
It is possible to formulate an Einstein
like theory with geometric/ physical objects depending additionally on
velocity type coordinates. The metric in such an endeavor  is taken to be $%
\mathbf{g}(x^{i},y^{a};y^{a_{1}},y^{a_{2}}),$ where $(y^{a_{1}},y^{a_{2}})$
are fiber coordinates. Such constructions were
originally elaborated upon in Finsler gravity \cite{vgrg}. The  motivation stems from
studying small  contributions from quantum gravity (QG) and MGTs that can be axiomatized on the
tangent bundle $TM_{1}^{3}$ in
phenomenological models with modified (non--quadratic) dispersion relations
(MDRs) see \cite{vaxiom}.


In order to construct generic off--diagonal solutions, it is convenient to
parameterize the local coordinates on $TM_{1}^{3}$ in the form $u^{\alpha
}=(x^{i},y^{a};y^{a_{1}},y^{a_{2}}),$ where $(x^{i},y^{a})$ are considered
for a conventional $2+2$ splitting on a base Lorentz manifold $M_{1}^{3}$
and $(y^{a_{1}},y^{a_{2}})$ are 2+2 coordinates for the typical fiber.
Indices correspondingly take  values $a_{1},b_{1},c_{1,}...=5,6$ and $%
a_{2},b_{2},c_{2},...=7,8.$  Possible dispersion relations of a MGT on $%
M_{1}^{3}$ are computed by perturbing the action of the theory around the
Minkowski background and using Fouries transforms of type $\psi (x^{\widehat{%
i}},t)=\int \frac{d^{3}p}{(2\pi )^{3/2}}\psi _{p}(t)e^{ip_{\widehat{i}}x^{%
\widehat{i}}},$  \cite{vgrg}.  For light rays propagation in a modified spacetime, the
MDR between the frequency $\omega $ and the wave vector $k_{\widehat{i}%
}\rightarrow p_{\widehat{i}}\sim y^{\widehat{i}}\,$ takes the
form
\begin{equation*}
\omega ^{2}(x^{i},y^{a};\varkappa ^{P})=c^{2}[g_{\widehat{i}\widehat{j}}y^{%
\widehat{i}}y^{\widehat{j}}]^{2}+[1-\frac{1}{r}q_{\widehat{i}_{1}\widehat{i}%
_{2}...\widehat{i}_{2r}}y^{\widehat{i}_{1}}y^{\widehat{i}_{2}}...y^{\widehat{%
i}_{2r}}(g_{\widehat{i}\widehat{j}}y^{\widehat{i}}y^{\widehat{j}})^{-2r}],
\end{equation*}%
where $c$ is the speed of light in vacuum and the coefficients $q_{\widehat{i%
}_{1}\widehat{i}_{2}...\widehat{i}_{2r}}$ are certain functions depending on space
coordinates and a set of parameters $\varkappa ^{P}$ defining a MGT model.
 If all $q_{\widehat{i}_{1}%
\widehat{i}_{2}...\widehat{i}_{2r}}$ are zero, we obtain the standard
quadratic dispersion relation for vacuum waves in special relativity, which holds
true at any point $(x^{i},y^{a})\in M_{1}^{3}.$

Let us associate to a MDR $\omega ^{2}(x^{i},y^{a};\varkappa ^{P})$ a
nonlinear quadratic element%
\begin{equation}
ds^{2}=F^{2}(x^{i},y^{a};y^{a_{1}},y^{a_{2}})\approx -(cdt)^{2}+g_{\widehat{i%
}\widehat{j}}y^{\widehat{i}}y^{\widehat{j}}[1+\frac{1}{r}q_{\widehat{i}_{1}%
\widehat{i}_{2}...\widehat{i}_{2r}}y^{\widehat{i}_{1}}y^{\widehat{i}%
_{2}}...y^{\widehat{i}_{2r}}(g_{\widehat{i}\widehat{j}}y^{\widehat{i}}y^{%
\widehat{j}})^{-r}]+\mathcal{O}(q^{2}).  \label{nfqe}
\end{equation}%
If there are satisfied the homogeneity conditions $F(x^{i},y^{a};\tau
y^{a_{1}},\tau y^{a_{2}})=\tau F(x^{i},y^{a};y^{a_{1}},y^{a_{2}})$ for any $%
\tau >1,$ and nondegenerate positive definite Hessian $g_{a_{z}b_{m}}=\frac{%
1}{2}\frac{\partial F^{2}}{\partial y^{a_{z}}\partial y^{b_{m}}};$ for $%
z,m=1,2,$ the value $F$ is termed  the Finsler fundamental function. For a more general class of theories with arbitrary frame
and coordinate transforms on $TM_{1}^{3},$ the condition of homogeneity of $%
F $ and positive definition of $g_{a_{z}b_{m}}$ are not usually imposed. We can
define as an effective Lagrangian $L=F^{2}$ on nonzero sections of $%
TM_{1}^{3}$ and express the corresponding Lagrange--Euler equations as
semi-spray equations,  \cite{vfinslbranes,vfinlbh,vacarunp1}.
This determines a canonical
nonlinear connection we call the  N--connection, the structure of  which is a Whitney sum $%
\mathbf{N}:TTM_{1}^{3}=hTM_{1}^{3}\oplus vTM_{1}^{3},$ for a nonholonomic
(equivalently, anholonomic, i.e nonintegrable) splitting into conventional
horizontal (h), and vertical (v) subspaces. Fixing a system of local
coordinates and frames with 2+2+2+2 splitting, a N--connection is determined by its
coefficients as {\small
\begin{equation*}
\mathbf{N}%
=\{N_{i}^{a}(x^{k},y^{b}),N_{i}^{a_{1}}(x^{k},y^{b},y^{b_{1}}),N_{i}^{a_{2}}(x^{k},y^{b},y^{b_{1}},y^{b_{2}}); N_{a}^{a_{1}}(x^{k},y^{b},y^{b_{1}}),N_{a}^{a_{2}}(x^{k},y^{b},y^{b_{1}},y^{b_{2}});
N_{a_{1}}^{a_{2}}(x^{k},y^{b},y^{b_{1}},y^{b_{2}})\}.
\end{equation*}%
\label{ncc} } We can associate to $\mathbf{N}$ a so--called N--adapted dual
basis {\small
\begin{equation}
\mathbf{e}^{\alpha }=(e^{i}=dx^{i},\mathbf{e}^{a}=dy^{a}+N_{i}^{a}dx^{i},%
\mathbf{e}^{a_{1}}=dy^{a_{1}}+N_{i}^{a_{1}}dx^{i}+N_{a}^{a_{1}}dy^{a},%
\mathbf{e}%
^{a_{2}}=dy^{a_{2}}+N_{i}^{a_{2}}dx^{i}+N_{a}^{a_{2}}dy^{a}+N_{a_{1}}^{a_{2}}dy^{a_{1}}).
\label{ddif}
\end{equation}%
}

Any metric $\mathbf{g}=\mathbf{g}_{\alpha \beta }(u^{\alpha })\mathbf{e}%
^{\alpha }\otimes \mathbf{e}^{\beta }$ can be written as a distinguished
metric (d--metric) with respect to a N--adapted coframe (\ref{ddif}),%
\begin{equation}
\mathbf{g}=g_{i}(x^{i})dx^{i}\otimes dx^{i}+g_{a}(x^{i},y^{b})\mathbf{e}%
^{a}\otimes \mathbf{e}^{a}+g_{a_{1}}(x^{i},y^{b},y^{b_{1}})\mathbf{e}%
^{a_{1}}\otimes \mathbf{e}^{a_{1}}+g_{a_{2}}(x^{i},y^{b},y^{b_{1}},y^{b_{2}})%
\mathbf{e}^{a_{2}}\otimes \mathbf{e}^{a_{2}}.  \label{dm}
\end{equation}%
As alluded to previously, one does not work in Finsler like gravity theories with the Levi--Civita
connection $\nabla $ because it is not adapted to the N--connection
splitting. Nevertheless, it is always possible to introduce an "auxiliary"
canonical distinguished connection (d--connection) $\widehat{\mathbf{D}}$
with a distortion relation $\widehat{\mathbf{D}}=\nabla +$ $\widehat{\mathbf{%
Z}}$ when both connections and the distorting tensor $\widehat{\mathbf{Z}}$
are uniquely determined by data $(\mathbf{g,N})$ following the conditions
that $\nabla \mathbf{g=0}$ with zero torsion $^{\nabla }\mathcal{T}$ of $%
\nabla $ and, respectively, $\widehat{\mathbf{D}}\mathbf{g}=0$ and the
"pure" $h$- and $v$--components of torsion $\widehat{\mathcal{T}}$ of $%
\widehat{\mathbf{D}}$ are zero. In general, there are nonzero $h$ and $v$
components of $\widehat{\mathcal{T}}$ \ but such values are induced by
anholonomy coefficients $W_{\beta \gamma }^{\alpha }[\mathbf{N}]$ for
\begin{equation}
\mathbf{e}_{\alpha }\mathbf{e}_{\beta }-\mathbf{e}_{\beta }\mathbf{e}%
_{\alpha }=W_{\alpha \beta }^{\gamma }\mathbf{e}_{\gamma }.  \label{anhrel}
\end{equation}%
The main idea of the so--called anholonomic frame deformation method (AFDM)
\cite{vkerrhd} is to consider $%
\widehat{\mathbf{D}}$ instead of $\nabla $ and find solutions of the
nonholonomically deformed Einstein equations%
\begin{equation}
\widehat{\mathbf{R}}_{ij}=\Lambda \mathbf{g}_{ij},\widehat{\mathbf{R}}%
_{ab}=\Lambda \mathbf{g}_{ab},\widehat{\mathbf{R}}_{a_{1}b_{1}}=\
^{1}\Lambda \mathbf{g}_{a_{1}b_{1}},\widehat{\mathbf{R}}_{a_{2}b_{2}}=\
^{2}\Lambda \mathbf{g}_{a_{2}b_{2}},  \label{densteq}
\end{equation}%
where $\widehat{\mathbf{R}}_{\alpha \beta }$ is the Ricci tensor of $%
\widehat{\mathbf{D}}$ and $\Lambda ,\ ^{1}\Lambda ,\ ^{2}\Lambda $ are
effective cosmological constants. The canonical d--connection allows us to decouple the
nonlinear system of partial differential equations, PDE, (\ref{densteq}) for
off--diagonal ansatz depending on all spacetime and extra--dimension
coordinates with respect to N--adapted frames (\ref{ddif}). We can integrate the decoupled system of PDE in very general forms
with Killing and non--Killing symmetries depending on generating and
integration functions and parameters, nontrivial effective sources from
extra dimensions and matter field interactions, or QG and MGTs terms. If needed, one can
impose additional (Levi--Civita, LC) constraints, $\widehat{\mathcal{T}}_{|%
\widehat{\mathbf{Z}}\rightarrow 0}\rightarrow 0,$ and extract
LC--configurations. Even for the zero torsion constraints, such solutions
are, in general, off--diagonal because the anholonomy coefficients $W_{\beta
\gamma }^{\alpha }[\mathbf{N}]$ are not zero. The fundamental priority of
the AFDM is that it admits exact solutions in GR and MGTs  with
nontrivial N--connection structure and generic off--diagonal interactions.

\subsection{Generic off--diagonal ansatz for stationary exact solutions}

In this subsection, we study the decoupling property of equations (\ref%
{densteq}) for metrics which are stationary in the 4D horizontal part and
with three Killing symmetries on $\partial _{3}=\partial _{t},\partial
_{5},\partial _{7}$ in the total space. We take the following parameterizations of the
N--adapted coefficients in (\ref{dm}) and (\ref{ddif}):
\begin{eqnarray}
g_{i} &=&\epsilon _{i}e^{q(x^{k})},\mbox{ for }\epsilon _{i}=\pm
1;g_{3}=h_{3}(x^{k},y^{4}),g_{4}=h_{4}(x^{k},y^{4});  \label{ansatz} \\
g_{5}
&=&h_{5}(x^{k},y^{4},y^{6}),g_{6}=h_{6}(x^{k},y^{4},y^{6});
g_{7}=h_{7}(x^{k},y^{4},y^{6},y^{8}),g_{8}=h_{8}(x^{k},y^{4},y^{6},y^{8}),
\notag \\
N_{i}^{3} &=&n_{i}(x^{k},y^{4}),N_{i}^{4}=w_{i}(x^{k},y^{4});  \notag \\
N_{i}^{5} &=&\ ^{1}n_{i}(x^{k},y^{4},y^{6}),N_{i}^{6}=\
^{1}w_{i}(x^{k},y^{4},y^{6});N_{a}^{5}=\
^{1}n_{a}(x^{k},y^{4},y^{6}),N_{a}^{6}=\ ^{1}w_{a}(x^{k},y^{4},y^{6});
\notag \\
N_{i}^{7} &=&\ ^{2}n_{i}(x^{k},y^{4},y^{6},y^{8}),N_{i}^{8}=\
^{2}w_{i}(x^{k},y^{4},y^{6},y^{8});N_{a}^{7}=\
^{2}n_{a}(x^{k},y^{4},y^{6},y^{8}),N_{a}^{8}=\
^{2}w_{a}(x^{k},y^{4},y^{6},y^{8});  \notag \\
N_{a_{1}}^{7} &=&\ ^{2}n_{a_{1}}(x^{k},y^{4},y^{6},y^{8}),N_{a_1}^{8}=\
^{2}w_{a_1}(x^{k},y^{4},y^{6},y^{8})\mbox{ for }i=1,2; a=3,4; a_{1}=5,6.
\notag
\end{eqnarray}%

Employing the ansatz (\ref{ansatz}) for the d--metric (\ref{dm}), the modified
Einstein equations (\ref{densteq}) transform into a system of decoupled PDE,
\begin{eqnarray}
\epsilon _{1}\partial _{1}^{2}q+\epsilon _{2}\partial _{2}^{2}q &=&2\Lambda ,
\label{sysnpde} \\
\partial _{4}\phi \ \partial _{4}h_{3} &=&2h_{3}h_{4}\Lambda ,\ \partial
_{4}^{2}n_{i}+\gamma \ \partial _{4}n_{i}=0,\ \beta w_{i}-\alpha _{i}=0,
\notag \\
\partial _{6}\ ^{1}\phi \ \partial _{6}h_{5} &=&2h_{5}h_{6}\ ^{1}\Lambda ,\
\partial _{6}^{2}\ ^{1}n_{i_{1}}+\ ^{1}\gamma \ \partial _{6}\
^{1}n_{i_{1}}=0,\ \ ^{1}\beta \ ^{1}w_{i_{1}}-\ ^{1}\alpha _{i_{1}}=0,
\notag \\
\partial _{8}\ ^{2}\phi \ \partial _{8}h_{7} &=&2h_{7}h_{8}\ ^{2}\Lambda ,\
\partial _{8}^{2}\ ^{2}n_{i_{2}}+\ ^{2}\gamma \ \partial _{8}\
^{2}n_{i_{2}}=0,\ \ ^{2}\beta \ ^{2}w_{i_{2}}-\ ^{2}\alpha _{i_{2}}=0.
\notag
\end{eqnarray}%
where the second partial derivatives are defined as $\partial _{1}^{2}a=\partial ^{2}a/\partial x^{1}\partial x^{1}$,
and similarly for the other indices  $i_{1}=(i,a),i_{2}=(i_{1},a_{1})$.
Additional LC--conditions can be imposed for zero torsion, leading to the following constraints%
\begin{eqnarray*}
\partial _{4}w_{i} &=&(\partial _{i}-w_{i}\partial _{4})\ln \sqrt{|h_{4}|}%
,(\partial _{i}-w_{i}\partial _{4})\ln \sqrt{|h_{3}|},\partial
_{i}w_{j}=\partial _{j}w_{i},\partial _{4}n_{i}=0; \\
\partial _{6}\ ^{1}w_{i_{1}} &=&(\partial _{i_{1}}-\ ^{1}w_{i_{1}}\partial
_{6})\ln \sqrt{|h_{6}|},(\partial _{i_{1}}-\ ^{1}w_{i_{1}}\partial _{6})\ln
\sqrt{|h_{5}|},\partial _{i_{1}}\ ^{1}w_{j_{1}}=\partial _{j_{1}}\
^{1}w_{i_{1}},\partial _{6}\ ^{1}n_{i_{1}}=0; \\
\partial _{8}\ ^{2}w_{i_{2}} &=&(\partial _{i_{2}}-\ ^{2}w_{i_{2}}\partial
_{8})\ln \sqrt{|h_{8}|},(\partial _{i_{2}}-\ ^{2}w_{i_{2}}\partial _{8})\ln
\sqrt{|h_{7}|},\partial _{i_{2}}\ ^{2}w_{j_{2}}=\partial _{j_{2}}\
^{2}w_{i_{2}},\partial _{8}\ ^{2}n_{i_{2}}=0.
\end{eqnarray*}%
The coefficients in above equations are given by formulas%
\begin{eqnarray*}
\phi &=&\ln |\frac{\partial _{4}h_{3}}{\sqrt{|h_{3}h_{4}|}}|,\gamma
=\partial _{4}(\ln \frac{|h_{3}|^{3/2}}{|h_{4}|}),\alpha _{i}=\frac{\partial
_{4}h_{3}}{2h_{3}}\partial _{i}\phi ,\beta =\frac{\partial _{4}h_{3}}{2h_{3}}%
\partial _{4}\phi , \\
\ ^{1}\phi &=&\ln |\frac{\partial _{6}h_{5}}{\sqrt{|h_{5}h_{6}|}}|,\
^{1}\gamma =\partial _{6}(\ln \frac{|h_{5}|^{3/2}}{|h_{6}|}),\ ^{1}\alpha
_{i_{1}}=\frac{\partial _{6}h_{5}}{2h_{5}}\partial _{i_{1}}\ ^{1}\phi ,\
^{1}\beta =\frac{\partial _{6}h_{5}}{2h_{5}}\partial _{6}\ ^{1}\phi , \\
\ ^{2}\phi &=&\ln |\frac{\partial _{8}h_{7}}{\sqrt{|h_{7}h_{8}|}}|,\
^{2}\gamma =\partial _{8}(\ln \frac{|h_{7}|^{3/2}}{|h_{8}|}),\ ^{2}\alpha
_{i_{2}}=\frac{\partial _{8}h_{7}}{2h_{7}}\partial _{i_{2}}\ ^{2}\phi ,\
^{2}\beta =\frac{\partial _{8}h_{7}}{2h_{7}}\partial _{8}\ ^{2}\phi ,
\end{eqnarray*}%
where
\begin{equation}
\partial _{4}h_{3}\neq 0,\partial _{6}h_{5}\neq 0,\partial _{8}h_{7}\neq 0%
\mbox{  if (respectively)  }\Lambda \neq 0,\ ^{1}\Lambda \neq 0\mbox{ and }\
^{2}\Lambda \neq 0.  \label{nondegen}
\end{equation}

We find exact solutions by integrating "step by step" the system (\ref%
{sysnpde}) for arbitrary generating functions $\phi (x^{k},y^{4}),$ $\
^{1}\phi (x^{k},y^{4},y^{6}),\ ^{2}\phi (x^{k},y^{4},y^{6},y^{8})$ (or $\Phi
=e^{\phi },\ ^{1}\Phi =e^{\ ^{1}\phi },\ ^{2}\Phi =e^{\ ^{2}\phi })$ and,
respectively, nonzero
 $\partial _{4}\phi ,\ \partial _{6}\ ^{1}\phi ,\ \partial _{8}\
^{2}\phi ;$ arbitrary integration functions $\ _{1}n_{i}(x^{k}),\
_{2}n_{i}(x^{k}),\ _{1}^{1}n_{i_{1}}(x^{k},y^{4}),\
_{2}^{1}n_{i_{1}}(x^{k},y^{4})$, $\ _{1}^{2}n_{i_{2}}(x^{k},y^{4},y^{6})$, $%
\ _{2}^{2}n_{i_{2}}(x^{k},y^{4},y^{6})$. Such generic off--diagonal
solutions are parameterized by quadratic elements%
\begin{eqnarray}
ds^{2} &=&e^{q(x^{k})}[\epsilon _{1}(dx^{1})^{2}+\epsilon _{2}(dx^{2})^{2}]+%
\frac{\Phi ^{2}}{4\Lambda }[dt+(\ _{1}n_{i}+\ _{2}n_{i}\int dy^{4}\frac{%
(\partial _{4}\Phi )^{2}}{\Phi ^{5}})dx^{i}]^{2}+\frac{(\partial _{4}\Phi
)^{2}}{\Lambda \Phi ^{2}}[dy^{4}+\frac{\partial _{i}\Phi }{\partial _{4}\Phi
}dx^{i}]^{2}  \notag \\
&&+\frac{(\ ^{1}\Phi )^{2}}{4\ ^{1}\Lambda }\left[ dy^{5}+\left( \
_{1}^{1}n_{i_{1}}+\ _{2}^{1}n_{i_{1}}\int dy^{6}\frac{(\partial _{6}\
^{1}\Phi )^{2}}{(\ ^{1}\Phi )^{5}}\right) dx^{i_{1}}\right] ^{2}+\frac{%
(\partial _{6}\ ^{1}\Phi )^{2}}{\ ^{1}\Lambda (\ ^{1}\Phi )^{2}}\left[
dy^{6}+\frac{\partial _{i_{1}}(\ ^{1}\Phi )}{\partial _{6}(\ ^{1}\Phi )}%
dx^{i_{1}}\right] ^{2}  \label{gensol} \\
&&+\frac{(\ ^{2}\Phi )^{2}}{4\ ^{2}\Lambda }\left[ dy^{7}+\left( \
_{1}^{2}n_{i_{2}}+\ _{2}^{2}n_{i_{2}}\int dy^{8}\frac{(\partial _{8}\
^{2}\Phi )^{2}}{(\ ^{2}\Phi )^{5}}\right) dx^{i_{2}}\right] ^{2}+\frac{%
(\partial _{6}\ ^{2}\Phi )^{2}}{\ ^{2}\Lambda (\ ^{2}\Phi )^{2}}\left[
dy^{8}+\frac{\partial _{i_{2}}(\ ^{2}\Phi )}{\partial _{8}(\ ^{2}\Phi )}%
dx^{i_{2}}\right] ^{2},  \notag
\end{eqnarray}%
for $x^{i_{2}}=(x^{i_{1}},y^{a_{1}})=(x^{i},y^{a},y^{a_{1}}).$ The solutions
(\ref{gensol}) are, in general, with nonholonomically induced torsion.


We extract pseudo--Riemannian metrics on $TM_{1}^{3}$ as solutions for
Einstein tangent bundles, i.e. for (\ref{densteq}) with $\widehat{\mathcal{T}%
}_{|\widehat{\mathbf{Z}}\rightarrow 0}\rightarrow 0$ if we
fix nonholonomic distributions subject to the LC--conditions of zero torsion when the
coefficients of the above d--metrics are subjected to additional conditions
and generated in the form%

\begin{eqnarray}
\Phi &=&\check{\Phi},\mbox{ for }\partial _{4}\partial _{i}\check{\Phi}%
=\partial _{i}\partial _{4}\check{\Phi};\ ^{1}\Phi =\ ^{1}\check{\Phi},%
\mbox{ for }\partial _{6}\partial _{i_{1}}\ ^{1}\check{\Phi}=\partial
_{i_{1}}\partial _{6}\ ^{1}\check{\Phi};\ ^{2}\Phi =\ ^{2}\check{\Phi},%
\mbox{ for }\partial _{2}\partial _{i_{2}}\ ^{2}\check{\Phi}=\partial
_{i_{2}}\partial _{8}\ ^{2}\check{\Phi};  \notag \\
\ _{2}n_{i} &=&0,\ _{2}^{1}n_{i_{1}}=0,\ _{2}^{2}n_{i_{2}}=0;\
_{1}n_{i}=\partial _{i}\ _{1}n(x^{k}),\ _{1}^{1}n_{i}=\partial _{i_{1}}\
_{1}^{1}n(x^{k},y^{4}),\ _{1}^{2}n_{i}=\partial _{i_{2}}\
_{1}^{2}n(x^{k},y^{4},y^{6});  \label{lccond} \\
w_{i} &=&\partial _{i}\check{\Phi}/\partial _{4}\check{\Phi}=\partial _{i}%
\check{A}(x^{k},y^{4}),  \notag \\
\ ^{1}w_{i_{1}} &=&\partial _{i_{1}}\ ^{1}\check{\Phi}/\partial _{6}\ ^{1}%
\check{\Phi}=\partial _{i_{1}}\ ^{1}\check{A}(x^{k},y^{4},y^{6}),\
^{2}w_{i_{2}}=\partial _{i_{2}}\ ^{2}\check{\Phi}/\partial _{8}\ ^{2}\check{%
\Phi}=\partial _{i_{2}}\ ^{2}\check{A}(x^{k},y^{4},y^{6},y^{8});  \notag \\
h_{3} &=&\check{\Phi}^{2}/4\Lambda ,h_{4}=(\partial _{4}\check{\Phi}%
)^{2}/\Lambda \check{\Phi}^{2},  \notag \\
h_{5} &=&\ ^{1}\check{\Phi}^{2}/4\ ^{1}\Lambda ,h_{6}=(\partial _{6}\ ^{1}%
\check{\Phi})^{2}/\ ^{1}\Lambda \ ^{1}\check{\Phi}^{2},\ h_{7}=\ ^{2}\check{%
\Phi}^{2}/4\ ^{2}\Lambda ,h_{8}=(\partial _{8}\ ^{2}\check{\Phi})^{2}/\
^{2}\Lambda \ ^{2}\check{\Phi}^{2};  \notag
\end{eqnarray}%
The generating and
integration functions in above formulas define integral varieties and
LC--subvarieties of generalized Einstein--Finsler spaces. Such functions can
be smooth , or with singular behaviour. For stationary
configurations and certain Killing symmetries, they can be determined from
asymptotic conditions for well--defined systems of coordinates and possible
limits to diagonal configurations with observational effects on the base
4D effective spacetime. We note that the AFDM allows us to construct
generic off--diagonal non--Killing solutions depending on all (higher
dimension) spacetime coordinates, \cite%
{vkerrhd,vfinslbranes,vfinlbh,vacarunp1}. In the next section, we elaborate on
the conditions when solutions of type (\ref{gensol}) with possible
LC--constraints  define black ring configurations with self--consistent
imbedding into velocity black ellipsoid/torus backgrounds.


The integral varieties of solutions determined by d--metrics of type (\ref%
{gensol}) (with possible constraints (\ref{lccond})) allows us to describe
nonholonomic deformations of locally anisotropic 8D spacetimes of "higher
symmetry" (for instance, with 2+2 Killing vectors) into configurations with
"lower symmetry" (for instance, with 1+1 Killing vectors). In such cases,
certain well defined physical models with higher symmetry can be extended to
less symmetric ones with certain conventional "polarization functions" and
effective interaction constants. In explicit form, we state these conditions
for N--adapted transforms
\begin{equation}
\lbrack \mathbf{\mathring{g}}_{\alpha }(x^{k},y^{a_{1}}),\mathbf{\mathring{N}%
}(x^{k},y^{a_{1}})]\mathbf{\rightarrow }[\mathbf{g}_{\alpha }=\eta _{\alpha
}(x^{k},y^{4};y^{a_{1}},y^{8})\mathbf{\mathring{g}}_{\alpha },\mathbf{N}=%
\mathbf{\mathring{N}+\check{N}}(x^{k},y^{4};y^{a_{1}},y^{8})],
\label{nonhdef}
\end{equation}%
where $\eta _{\alpha }$ and $\mathbf{\check{N}}$ are respective
"gravitational polarizations" of the coefficients of d--metric and
N--connection structures.\footnote{%
We do not consider summation on indices for values of type $\eta _{\alpha }%
\mathbf{\mathring{g}}_{\alpha }$} Applying the AFDM, we construct
nonholonomic deformations when the "prime" data $[\mathbf{\mathring{g}}%
_{\alpha },\mathbf{\mathring{N}}]$ may be, or not, a solution of some
gravitational field equations but, positively, the "target" data $[\mathbf{g}%
_{\alpha }=\eta _{\alpha }\mathbf{\mathring{g}}_{\alpha },\mathbf{N}=\mathbf{%
\mathring{N}+\check{N}}]$ are determined by some exact solutions of
generalized Einstein--Finsler equations.

For certain physically important cases, we can consider that $[\mathbf{%
\mathring{g}}_{\alpha },\mathbf{\mathring{N}}]$ are just certain trivial
lifts from a 4D stationary vacuum spacetime with two Killing symmetries,
when $\Lambda =\ ^{1}\Lambda =\ ^{2}\Lambda =0.$ The target solutions $\left[
\mathbf{g}_{\alpha },\mathbf{N}\right] $ for an ansatz (\ref{ansatz}) and a
d--metric (\ref{dm}) belong, in general, to an integral variety with nonzero
effective cosmological constants $\Lambda ,\ ^{1}\Lambda ,\ ^{2}\Lambda $
(up to certain classes of frame transforms and re--definitions of generating
functions and effective sources of gravitational field equations). We can
provide similar physical interpretations both for the prime and target
metrics in a nonholonomic deformation
\begin{equation}
\lbrack \mathbf{\mathring{g}}_{\alpha },\mathbf{\mathring{N}}]\rightarrow
\lbrack \mathbf{g}_{\alpha }(\varepsilon )=\eta _{\alpha }\mathbf{\mathring{g%
}}_{\alpha },\ \mathbf{N}(\varepsilon )=\mathbf{\mathring{N}+\ \check{N}}%
(\varepsilon )],\mbox{
with }\eta _{\alpha }=1+\varepsilon \chi _{\alpha }\mbox{ and }\mathbf{\
\check{N}}(\varepsilon )=\varepsilon \ \mathbf{\check{N},}  \label{epsdef}
\end{equation}%
for a small parameter $0<\varepsilon \ll 1.$ Such nonholonomic transforms
may not have a smooth $\lim_{\varepsilon \rightarrow 0}\mathbf{g}%
(\varepsilon )\rightarrow \mathbf{\mathring{g}}$ if, for instance, $\mathbf{%
\mathring{g}}$ is a solution of vacuum Einstein equations (with possible
trivial extensions on total space $TM_{1}^{3}$) but $\mathbf{g}(\varepsilon
) $ is a soluton of MGTs with nontrivial sources. In general, $[\mathbf{%
\mathring{g}}_{\alpha },\mathbf{\mathring{N}}]$ and $[\mathbf{g}_{\alpha
}(\varepsilon ),\mathbf{N}(\varepsilon )]$ may describe nonholonomic
configurations with different topology and symmetries and/or very different
geometric/ physical models. There are necessary additional analysis of the
properties of nonholonomic deformations of certain prime geometric data into
nonholonomic ones. For small values $\varepsilon ,$ it is possible to state
such conditions on generating and integration functions and, for instance,
for effective de Sitter configurations on $TM_{1}^{3},$ when two
configurations may have analogous behaviour but with certain deformed
symmetries and effective nonlinear "polarization" functions.

\section{ Black Ring and Kerr Solutions}

\label{s3}The black ring and black hole solutions constructed and reviewed
in \cite{emp,emp1} occur in extra dimension spacetimes, for instance, in
(super) string / gravity and Kaluza--Klein theories. In this
section we show how such solutions can be constructed on $TM_{1}^{3}.$
These constructions have a different physical interpretation because
higher dimension coordinates are  velocity type coordinates and generalize
on respective tangent bundles certain 4D gravity theories, in particular,
the Einstein gravity.

\subsection{Neutral black ring solutions on tangent Lorentz bundle}

We introduce 5D ring coordinates as in sections 2-3 of \cite{emp1} and
extend them trivially to 8D, $x^{1}(x),x^{2}(y),y^{3}=t,y^{4}=\chi
,y^{5}=\psi ,y^{6},y^{7},y^{8}$, where
\begin{equation*}
x^{1}(x)=\int dx|G(x)|^{-1/2}\mbox{ and }x^{2}(y)=\int dy|G(y)|^{-1/2},%
\mbox{ for }G(\xi )=(1-\xi ^{2})(1+\nu \xi ),
\end{equation*}%
and the constants $\lambda ,\nu ,\mathring{c}$ and function $F(\xi
)=1+\lambda \xi $ are related via $\mathring{c}:=\sqrt{\lambda (\lambda -\nu
)(1+\lambda )/(1-\lambda )}$ in order to get black ring configurations for $%
0<\nu \leq \lambda <1.$\footnote{%
In this paper, we write $\chi $ instead the angle variable $\phi $ in R.
Emparan,\ H. S. Real and others' works because in our approach the symbol $%
\phi $ is used for generating functions (we can construct in similar forms
alternative classes of solutions with $y^{5}=\chi ,y^{4}=\psi )$ and work
with inverse functions $x(x^{1})$ and $y(x^{2})$ to the respective ones
defined above.} The heuristic construction of a 5D black ring is $%
TM_{1}^{3}$ can be considered as a boosted black string with velocity like
coordinate $\psi $ (on such strings, see \cite{vacarunp1} and references
therein) bent into a circular shape and imbedded into a total space with
curved fiber subspace determined by a velocity black hole configuration. The
nontrivial coefficients for a "prime" metric $\mathbf{\mathring{g}}_{\alpha
\beta }$ are chosen
\begin{eqnarray}
\mathring{g}_{1}(x^{k}) &=&e^{\mathring{g}(x^{1},x^{2})},\mathring{g}%
_{2}(x^{k})=-e^{\mathring{g}(x^{1},x^{2})},\mathring{h}_{3}(x^{k})=A+\frac{%
(AB)^{2}}{AB^{2}-S},\mathring{h}_{4}(x^{k})=\frac{\mathring{r}^{2}G(x^{1})}{%
[x(x^{1})-y(x^{2})]^{2}},  \notag \\
\mathring{h}_{5}(x^{k}) &=&AB^{2}-S,\mathring{h}_{6}=1,\mathring{h}_{7}=\pm
1,\mathring{h}_{8}=1,\mathring{N}_{3}^{5}=\ ^{1}\mathring{w}_{3}(x^{k})=%
\frac{AB}{AB^{2}-S},  \label{data2}
\end{eqnarray}%
for $A=-F(x^{1})/F(x^{2}),B=\mathring{c}\mathring{r}(1+y(x^{2}))/F(x^{2}),S=-%
\mathring{r}^{2}F(x^{1})G(x^{2})/F(x^{2})[x(x^{1})-y(x^{2})]^{2},$ where $%
\mathring{r}=const$ ($\mathring{r}$ is used instead of the radial constant $%
R $ in \cite{emp1}; see there the section 3.1 related to the vacuum solution
of 5-d Einstein equations by formula (14)). The function $%
\mathring{g}(x^{1},x^{2})$ is defined from the relation
\begin{eqnarray*}
e^{\mathring{g}}[(dx^{1})^{2}-(dx^{2})^{2}] &=&\frac{\mathring{r}^{2}}{%
(x-y)^{2}}F(x)[\frac{dx^{2}}{G(x)}-\frac{dy^{2}}{F(y)}] \\
\mbox{ and }A(dt-Bd\psi )^{2}-Sd\psi ^{2} &=&\mathring{h}_{3}(x^{k})dt^{2}+%
\mathring{h}_{5}(x^{k})[d\psi +\ ^{1}\mathring{w}_{3}(x^{k})dt]^{2}.
\end{eqnarray*}%
We obtain an example of d--metric (\ref{ddif}) when the coefficients depend
 only on two coordinates $x^{k}(x,y),$
\begin{equation}
ds^{2}=e^{\mathring{g}(x^{k})}[(dx^{1})^{2}-(dx^{2})^{2}]+\mathring{h}%
_{3}(x^{k})dt^{2}+\mathring{h}_{4}(x^{k})d\chi ^{2}+(dy^{5})^{2}\pm
(dy^{6})^{2}\pm (dy^{7})^{2}+\mathring{h}_{5}(x^{k})[d\psi +\ ^{1}\mathring{w%
}_{3}(x^{k})dt]^{2},  \label{blring}
\end{equation}%
where we have changed $y^{5}$ into $y^{8}=\psi $ (this is convenient for
further applications of the AFDM). This metric is mathematically equivalent
to that for the neutral black ring solution with trivial extension from 5D
to 8D in variables on $TM_{1}^{3}.$ The sign $\pm $ before $dy^{6}$ and/or
$dy^{7}$ reflects two possibilities to lift geometric objects on $M_{1}^{3}$
in the total space. Nevertheless, this is not a model of "two time" physics
\cite{twotime1,twotime2,twotime3} if the geometric/ physical objects on $%
TM_{1}^{3}$ are considered as certain canonical lifts (for instance, of
Sasaki type \cite{yano}) from $M_{1}^{3}$ with one time like coordinate. In
our approach, the extra dimensional coordinate $y^{8}=\psi $ is of velocity
type, i.e. our analogous black ring solution is for a model of "phase" space
on tangent Lorentz bundle. We conclude that the class of solutions (\ref%
{blring}) constitute an example of metrics of type (\ref{gensol}) which are
degenerate in the sense that the conditions (\ref{nondegen}) are not
satisfied. This is a possibility if $\partial _{4}\mathring{h}_{4}=0$ and $%
\Lambda =0$ for vacuum solutions.

\subsection{Black ring solutions in velocity Kerr backgrounds}

We provide an example when the metric (\ref{blring}) is nonholonomically
deformed by corresponding MDR (up to frame/coordinate transforms on $
TM_{1}^{3}$ and/or $M_{1}^{3})$ into an analogous black ring interacting
with an analogous Kerr black hole determined by velocity type coordinates \cite{heusler,kramer,misner}.  We use
$\psi$ instead of $\varphi $ for respective fiber's signature $(+,-,+,+).$ Also, we
work with correspondingly N--adapted frames and systems of coordinates
instead of the "standard" prolate spherical, or the  Boyer--Linquist coordinates. The fiber analogs of Boyer--Linquist coordinates $(\widetilde{y}%
^{a_{1}^{\prime }};\widetilde{y}^{a_{2}^{\prime }}),$ for $a_{1}^{\prime
}=5,6$ and $a_{2}^{\prime }=7,8,$ are defined $\widetilde{y}^{5}=\widetilde{t%
},\widetilde{y}^{6}=\widetilde{y}^{6}(\widetilde{r},\widetilde{\vartheta }),%
\widetilde{y}^{7}=\widetilde{y}^{7}(\widetilde{r},\widetilde{\vartheta }),%
\widetilde{y}^{8}=\psi =y^{8},$ where "tilde" \ is used in order to
emphasize that such coordinates are analogous to the ones on a fiber space. The
total black hole analogous mass is $m_{0},$ and the analogous total angular
momentum is $am_{0},$ for the asymptotically flat, stationary and
axisymmetric Kerr in the space of relativistic velocities. In such
variables, the 8D vacuum metric (\ref{blring}) is generalized TO the form
\begin{eqnarray}
ds^{2} &=&e^{\mathring{g}}[(dx^{1})^{2}-(dx^{2})^{2}]+\mathring{h}_{3}dt^{2}+%
\mathring{h}_{4}d\chi ^{2}+(\widetilde{A}-\widetilde{B}^{2}/\widetilde{C})(d%
\widetilde{t})^{2}+\widetilde{\Xi }\widetilde{\Delta }^{-1}(d\widetilde{r}%
)^{2}+\widetilde{\Xi }(d\widetilde{\vartheta })^{2}  \notag \\
&&+\mathring{h}_{5}\widetilde{C}[d\psi +\ ^{1}\mathring{w}_{3}(x^{k})dt+%
\widetilde{B}^{2}/\widetilde{C}d\widetilde{t}]^{2}  \notag \\
&=&\mathring{g}_{k}dx^{k}+\mathring{h}_{3}(x^{i})dt^{2}+\mathring{h}%
_{4}(x^{i})d\chi ^{2}+\widetilde{h}_{5}(\widetilde{y}^{6},\widetilde{y}^{7})d%
\widetilde{t}^{2}+\widetilde{h}_{6}(\widetilde{y}^{6},\widetilde{y}^{7})(d%
\widetilde{y}^{6})^{2}  \notag \\
&&+\widetilde{h}_{7}(\widetilde{y}^{6},\widetilde{y}^{7})(d\widetilde{y}%
^{7})^{2}+\widetilde{h}_{8}(x^{k},\widetilde{y}^{6},\widetilde{y}^{7})[d\psi
+\widetilde{N}_{3}^{8}dt+\widetilde{N}_{5}^{8}d\widetilde{t}]^{2},
\label{blrblkerrbl}
\end{eqnarray}%
where $\widetilde{\Xi }[\widetilde{\Delta }^{-1}(d\widetilde{r})^{2}+(d%
\widetilde{\vartheta })^{2}]=\widetilde{h}_{6}(d\widetilde{y}^{6})^{2}+%
\widetilde{h}_{7}(\widetilde{y}^{6},\widetilde{y}^{7})(d\widetilde{y}%
^{7})^{2}$ for certain coefficients $\widetilde{h}_{6}(\widetilde{y}^{6},%
\widetilde{y}^{7})$ and $\widetilde{h}_{7}(\widetilde{y}^{6},\widetilde{y}%
^{7}),$ $\mathring{h}_{5}(x^{k})=AB^{2}-S,$ and
\begin{eqnarray}
\widetilde{A}(\widetilde{r},\widetilde{\vartheta }) &=&-\widetilde{\Xi }%
^{-1}(\widetilde{\Delta }-a^{2}\sin ^{2}\widetilde{\vartheta }),\widetilde{B}%
=\widetilde{\Xi }^{-1}a\sin ^{2}\vartheta \left[ \widetilde{\Delta }-(%
\widetilde{r}^{2}+a^{2})\right] ,  \label{kerrcoef} \\
\widetilde{C}(\widetilde{r},\widetilde{\vartheta }) &=&\widetilde{\Xi }%
^{-1}\sin ^{2}\widetilde{\vartheta }\left[ (\widetilde{r}^{2}+a^{2})^{2}-%
\widetilde{\Delta }a^{2}\sin ^{2}\widetilde{\vartheta }\right] ,\mbox{ and }%
\widetilde{\Delta }(\widetilde{r}^{2})=\widetilde{r}^{2}-2m_{0}+a^{2},\
\widetilde{\Xi }(\widetilde{r},\widetilde{\vartheta })=\widetilde{r}%
^{2}+a^{2}\cos ^{2}\widetilde{\vartheta }.  \notag
\end{eqnarray}

The nonlinear quadratic line element (\ref{blrblkerrbl}) is determined by
data
\begin{eqnarray}
\mathring{g}_{1} &=&e^{\mathring{g}(x^{k})},\mathring{g}_{2}=-e^{\mathring{g}%
(x^{k})},\mathring{h}_{3}=A(x^{k})+\frac{[A(x^{k})B(x^{k})]^{2}}{%
A(x^{k})B^{2}(x^{k})-S(x^{k})},\mathring{h}_{4}=\frac{\mathring{r}%
^{2}G(x^{1})}{[x(x^{1})-y(x^{2})]^{2}},  \label{dkerr} \\
\widetilde{h}_{5} &=&\widetilde{A}-\widetilde{B}^{2}/\widetilde{C},%
\widetilde{h}_{6}=\widetilde{h}_{6}(\widetilde{y}^{6},\widetilde{y}^{7}),%
\widetilde{h}_{7}=\widetilde{\Xi },\widetilde{h}_{8}=\mathring{h}_{5}%
\widetilde{C},\widetilde{N}_{3}^{8}=\ ^{1}\mathring{w}_{3}(x^{k}),\widetilde{%
N}_{5}^{8}=\ ^{2}\mathring{n}_{5}=\widetilde{B}^{2}/\widetilde{C},  \notag
\end{eqnarray}%
and defines solutions of vacuum Einstein equations on $TM_{1}^{3}.$ For any
fixed point $(x^{k},t,\chi )\in M_{1}^{3},$ the fiber coefficients (\ref%
{dkerr}) of metric in velocity type fiber variables $(\widetilde{t},%
\widetilde{r},\widetilde{\vartheta },\psi )$ determine an analogous Kerr
like black hole with effective time like coordinate $\widetilde{y}^{5}=%
\widetilde{t}.$ The black ring configuration is self--consistently embedded
in the velocity subspace via terms $\widetilde{\mathring{h}}_{8}=\mathring{h}%
_{5}\widetilde{C}$ and $\widetilde{N}_{3}^{8}=\ ^{1}\mathring{w}_{3}(x^{k}).$

\section{Off--diagonally Deformed Black Rings and Kerr--de Sitter Metrics}

\label{s4} For a nonvanishing cosmological constant, the vacuum metrics (\ref%
{blrblkerrbl}) can be nonholonomically deformed into exact solutions of type
(\ref{gensol}) for the gravitational field equations (\ref{densteq}). There are two varieties of solutions, the stationary ones exhibiting Killing symmetry in $\partial /\partial t$ and the ones on solitonic backgrounds that depend  explicitly on time. We present representative examples of both in the following.
We consider N--adapted deformations (\ref{nonhdef}) of the prime data (\ref%
{dkerr}), $\ \mathbf{\mathring{g}}_{\alpha \beta }=[\mathring{g}_{i},%
\mathring{h}_{a},\widetilde{h}_{a_{1}},\widetilde{h}_{a_{2}},\widetilde{N}%
_{3}^{8},\widetilde{N}_{6}^{8};\Lambda =\ ^{1}\Lambda =\ ^{2}\Lambda =0],$
into target data for an ansatz (\ref{ansatz}) and d--metric (\ref{dm}),
\begin{eqnarray*}
\mathbf{g}_{\alpha \beta } &=&[\eta _{i}\mathring{g}_{i},\eta _{a}\mathring{h%
}_{a},\eta _{a_{1}}\widetilde{h}_{a_{1}},\eta _{a_{2}}\widetilde{h}%
_{a_{2}};N_{i}^{a}=\check{N}_{i}^{a},N_{i}^{a_{1}}=\check{N}%
_{i}^{a_{1}},N_{i}^{a_{2}}=\delta _{8}^{a_{2}}\delta _{i}^{3}\widetilde{N}%
_{3}^{8}+\check{N}_{i}^{a_{2}}; \\
&&N_{a}^{a_{1}}=\check{N}_{a}^{a_{1}},N_{a}^{a_{2}}=\check{N}%
_{a}^{a_{2}};N_{a_{1}}^{a_{2}}=\delta _{8}^{a_{2}}\delta _{a_{1}}^{5}%
\widetilde{N}_{5}^{8}+\check{N}_{a_{1}}^{a_{2}};\Lambda \neq 0,\ ^{1}\Lambda
\neq 0,\ ^{2}\Lambda \neq 0],
\end{eqnarray*}%
when vacuum 8D solutions in EFG are transformed into off--diagonal
 de Sitter configurations on $TM_{1}^{3}$.

\subsection{Stationary solutions}
(A.) Off--diagonal deformations of the vacuum solutions  (\ref{blrblkerrbl}) are described by target d--metrics
\begin{eqnarray}
ds^{2} &=&e^{\mathring{g}}[(dx^{1})^{2}-(dx^{2})^{2}]+\eta _{3}\mathring{h}%
_{3}(dt+\check{n}_{i}dx^{i})^{2}+\eta _{4}\mathring{h}_{4}(d\chi +\check{w}%
_{i}dx^{i})^{2}  \label{abrkerrbackgr} \\
&&+\eta _{5}(\widetilde{A}-\widetilde{B}^{2}/\widetilde{C})(d\widetilde{t}+\
^{1}\check{n}_{i_{1}}dx^{i_{1}})^{2}+\eta _{6}\widetilde{\Xi }\widetilde{%
\Delta }^{-1}(d\widetilde{r}+\ ^{1}\check{w}_{i_{1}}dx^{i_{1}})^{2}  \notag
\\
&&+\eta _{7}\widetilde{\Xi }(d\widetilde{\vartheta }+\ ^{2}\check{n}%
_{i_{2}}dx^{i_{2}})^{2}+\eta _{8}\mathring{h}_{5}\widetilde{C}[d\psi +\ ^{2}%
\check{w}_{i_{2}}dx^{i_{2}}+\ ^{1}\mathring{w}_{3}(x^{k})dt+(\widetilde{B}%
^{2}/\widetilde{C})d\widetilde{t}]^{2},  \notag
\end{eqnarray}%
for $x^{i_{2}}=(x^{1}(x),x^{2}(x),y^{3}=t,y^{4}=\chi ,\widetilde{y}^{5}=%
\widetilde{t},\widetilde{y}^{6}=\widetilde{r},\widetilde{y}^{7}=\widetilde{%
\vartheta },\widetilde{y}^{8}=\psi ).$ The $\eta $--polarizations and
N--coefficients, respectively, are determined in this form:
\begin{eqnarray}
&&\eta _{i}\text{ \mbox{ from } }e^{\mathring{g}}\eta _{1}=e^{\mathring{g}%
}\eta _{2}=e^{q},\mbox{ for }q(x^{k})%
\mbox{ being  solution of the first
equation in (\ref{sysnpde}) };  \label{data1} \\
&&\eta _{a}\mbox{  computed as }\eta _{3}=\Phi ^{2}/4\Lambda \mathring{h}_{3}%
\mbox{ and }\eta _{4}=(\partial _{4}\Phi )^{2}/\Lambda \Phi ^{2}\mathring{h}%
_{4},\ \forall \ \Phi (x^{k},y^{4}),\partial _{4}\Phi \neq 0;  \notag \\
&&\eta _{a_{1}}\mbox{ as }\eta _{5}=(\ ^{1}\Phi )^{2}\widetilde{C}/4\
^{1}\Lambda (\widetilde{C}\widetilde{A}-\widetilde{B}^{2})\mbox{ and }\eta
_{6}=(\partial _{6}\ ^{1}\Phi )^{2}\widetilde{\Delta }/\ ^{1}\Lambda (\
^{1}\Phi )^{2}\widetilde{\Xi },\ \forall \ \ ^{1}\Phi
(x^{k},y^{4},y^{6}),\partial _{6}\ ^{1}\Phi \neq 0;  \notag \\
&&\eta _{a_{2}}\mbox{  as }\eta _{7}=(\ ^{2}\Phi )^{2}/4\ ^{2}\Lambda
\widetilde{\Xi }\mbox{ and }\eta _{8}=(\partial _{8}\ ^{2}\Phi )^{2}/\
^{2}\Lambda (\ ^{2}\Phi )^{2}\mathring{h}_{5}\widetilde{C},\ \forall \ \
^{2}\Phi (x^{k},y^{4},y^{6},y^{8}),\partial _{8}\ ^{2}\Phi \neq 0,
\mbox{\
and \ }  \notag
\end{eqnarray}%
\begin{eqnarray}
N_{i}^{3} &=&\check{N}_{i}^{3}=\check{n}_{i}(x^{k},y^{4})=\ _{1}\check{n}%
_{i}(x^{k})+\ _{2}\check{n}_{i}(x^{k})\int d\chi (\partial _{4}\Phi
)^{2}/\Phi ^{5},\ N_{i}^{4}=\check{N}_{i}^{4}=\check{w}_{i}(x^{k},y^{4})=%
\partial _{i}\Phi /\partial _{4}\Phi ;  \notag \\
N_{i_{1}}^{5} &=&\check{N}_{i_{1}}^{5}=\ ^{1}\check{n}%
_{i_{1}}(x^{k},y^{4},y^{6})=\ _{1}^{1}\check{n}_{i_{1}}(x^{k},y^{4})+\
_{2}^{1}\check{n}_{i_{1}}(x^{k},y^{4})\int d\widetilde{r}(\partial _{6}\
^{1}\Phi )^{2}/(\ ^{1}\Phi )^{5},  \notag \\
N_{i_{1}}^{6} &=&\check{N}_{i_{1}}^{6}=\ ^{1}\check{w}%
_{i_{1}}(x^{k},y^{4},y^{6})=\partial _{i_{1}}\ ^{1}\Phi /\partial _{6}\
^{1}\Phi ;  \label{data2a} \\
N_{i_{2}}^{7} &=&\check{N}_{i_{2}}^{7}=\ ^{2}\check{n}%
_{i_{2}}(x^{k},y^{4},y^{6},y^{8})=\ _{1}^{2}\check{n}%
_{i_{2}}(x^{k},y^{4},y^{6})+\ _{2}^{2}\check{n}_{i_{2}}(x^{k},y^{4},y^{6})%
\int d\psi (\partial _{8}\ ^{2}\Phi )^{2}/(\ ^{2}\Phi )^{5},  \notag \\
N_{i_{2}}^{8} &=&\delta _{i_{2}}^{3}\widetilde{N}_{3}^{8}+\delta _{i_{2}}^{5}%
\widetilde{N}_{5}^{8}+\check{N}_{i_{2}}^{8}=\delta _{i_{2}}^{3}\ ^{1}%
\mathring{w}_{3}+\delta _{i_{2}}^{5}(\widetilde{B}^{2}/\widetilde{C})+\ ^{2}%
\check{w}_{i_{2}}(x^{k},y^{4},y^{6})=\partial _{i_{2}}\ ^{2}\Phi /\partial
_{8}\ ^{2}\Phi ;  \notag
\end{eqnarray}%
for $i_{1}=(i,a)=1,2,3,4$ and $i_{2}=(i,a,a_{1})=1,2,3,4,5,6,$ where we
fixed the integration functions to get a N--adapted $2+2+2+2$ splitting.

Geometrically, d--metrics (\ref{abrkerrbackgr}) describe nonholonomic
imbedding of analogous black ring and velocity Kerr metrics into more
general off-diagonal configurations determined by generation functions
(values $\Phi ,$ $\ ^{1}\Phi $ and $\ ^{2}\Phi $) and integration functions
of type $\ _{1}\check{n}_{i},\ _{2}\check{n}_{i},$ $\ _{1}^{1}\check{n}%
_{i_{1}},\ _{2}^{1}\check{n}_{i_{1}}$ etc. The target metrics \ are with
less symmetries but still contain Killing symmetries on $\partial
_{3},\partial _{5}$ and $\partial _{7}.$ We can compute nontrivial torsion
components following formulas $\widehat{\mathcal{T}}^{\alpha }:=\ \widehat{%
\mathbf{D}}\mathbf{e}^{\alpha }=d\mathbf{e}^{\alpha }+\widehat{\mathbf{%
\Gamma }}_{\ \beta }^{\alpha }\wedge \mathbf{e}^{\beta }$. The coefficients
of 1--form $\widehat{\mathbf{\Gamma }}_{\ \beta }^{\alpha }=\widehat{\mathbf{%
\Gamma }}_{\ \beta \gamma }^{\alpha }\mathbf{e}^{\gamma }$ are determined by
the canonical d--connection $\ \widehat{\mathbf{D}}.$ Such a torsion
characterises a nonholonomic structure with nontrivial anholonomy
coefficients $W_{\beta \gamma }^{\alpha }$ in (\ref{anhrel}). It is
different from the torsion fields, for instance, in the Einstein--Cartan or
string gravity theories because it is not determined by additional dynamical
degrees of freedom and algebraic/ dynamical sources. The terms with
"up-circle" label like $\mathring{h}_{a},$ $e^{\mathring{g}}$ etc encode
prime data for analogous black ring configurations. We also distinguish
prime Kerr data by values tilde, like $\widetilde{C},\widetilde{A}$ and $%
\widetilde{B}.$ The term $\ ^{1}\mathring{w}_{3}(x^{k})$ for $\eta
_{8}\rightarrow 1$ describe interactions between black ring and velocity
black holes. A physical interpretation for
generic off--diagonal solutions  is lacking for  general nonholonomic deformations with arbitrary generating
and integration functions.

\vspace{0.5 in}
\noindent (B.) We can prescribe  nonholonomic distributions with N--adapted frame
structures on $TM_{1}^{3}$ when the condition $\widehat{\mathcal{T}}^{\alpha
}=0$ is satisfied for $\Phi =\check{\Phi},\ ^{1}\Phi =\ ^{1}\check{\Phi},\
^{2}\Phi =\ ^{2}\check{\Phi}$ and the integration functions are subjected to
the conditions (\ref{lccond}). Introducing such values in (\ref{data1}) and (%
\ref{data2a}), we generate LC--configurations with effective quadratic
element {\small
\begin{eqnarray}
ds^{2} &=&e^{q}[\epsilon _{1}(dx^{1})^{2}+\epsilon _{2}(dx^{2})^{2}]+\frac{%
\check{\Phi}^{2}}{4\Lambda }[dt+\partial _{i}(\ _{1}n)dx^{i}]^{2}+\frac{%
(\partial _{4}\check{\Phi})^{2}}{\Lambda \check{\Phi}^{2}}[d\chi +(\partial
_{i}\check{A})dx^{i}]^{2} +\frac{(\ ^{1}\check{\Phi})^{2}}{4\ ^{1}\Lambda }[d%
\widetilde{t}+(\partial _{i_{1}}\ _{1}^{1}n)dx^{i_{1}}]^{2}  \notag \\
&& + \frac{(\partial _{6}\ ^{1}\check{\Phi})^{2}}{\ ^{1}\Lambda (\ ^{1}%
\check{\Phi})^{2}}[d\widetilde{r}+(\partial _{i_{1}}\ ^{1}\check{A}%
)dx^{i_{1}}]^{2} +\frac{(\ ^{2}\check{\Phi})^{2}}{4\ ^{2}\Lambda }[d%
\widetilde{\vartheta }+(\partial _{i_{2}}\ _{1}^{2}n)dx^{i_{2}}]^{2}+\frac{%
(\partial _{8}\ ^{2}\check{\Phi})^{2}}{\ ^{2}\Lambda (\ ^{2}\check{\Phi})^{2}%
}[d\psi +(\partial _{i_{2}}\ ^{2}\check{A})dx^{i_{2}}]^{2}.
\label{abrkerrbackgr1}
\end{eqnarray}%
}  Metrics of this type define generic off--diagonal Einstein manifolds with nonzero
effective cosmological constant and possible polarization in the fiber space
and velocity like variables generated by nonholonomic deformations of the
prime metrics (\ref{blrblkerrbl}). For small polarizations $\eta _{\alpha
}=1+\varepsilon \chi _{\alpha }$ and$\mathbf{\ \check{N}}(\varepsilon
)=\varepsilon \ \mathbf{\check{N}}$ (\ref{epsdef}), we model analogous black
ring and black holes with velocity type variables imbedded into
asymptotically flat configurations on $TM_{1}^{3}.$ In general, the
solutions (\ref{abrkerrbackgr1}) can not be diagonalized for coordinate
transforms if the anholonomy coefficients $W_{\beta \gamma }^{\alpha }$ (\ref%
{anhrel}) are not zero and the N--connection coefficients are not trivial for
MDRs.

\label{ssbrbe}Schwarzschild configurations can be extracted from the Kerr
ones if $a=0$ in (\ref{kerrcoef}). We denote
\begin{equation}
\check{A}(\widetilde{r})=\widetilde{A}_{|a=0}=-\widetilde{r}^{-2}\check{%
\Delta},\check{B}=\tilde{B}_{|a=0}=0,\check{C}(\widetilde{r},\widetilde{%
\vartheta })=\widetilde{C}_{|a=0}=\widetilde{r}^{2}\sin ^{2}\widetilde{%
\vartheta },\mbox{ and }\check{\Delta}(\widetilde{r}^{2})=\widetilde{r}%
^{2}-2m_{0},\check{\Xi}=\widetilde{r}^{2}.  \notag
\end{equation}%
Considering polarization functions with a small parameter $\varepsilon ,$ we
construct the off--diagonal solutions
\begin{eqnarray}
ds^{2} &=&e^{\mathring{g}}[(dx^{1})^{2}-(dx^{2})^{2}]+(1+\varepsilon \chi
_{3})\mathring{h}_{3}\left[ dt+\varepsilon \partial _{i}(\ _{1}n)dx^{i}%
\right] ^{2}+(1+\varepsilon \chi _{4})\mathring{h}_{4}\left[ d\chi
+\varepsilon (\partial _{i}\check{A})dx^{i}\right] ^{2}  \notag \\
&&+(1+\varepsilon \chi _{5})\check{A}\left[ d\widetilde{t}+\varepsilon
(\partial _{i_{1}}\ _{1}^{1}n)dx^{i_{1}}\right] ^{2}+(1+\varepsilon \chi
_{6})\widetilde{r}^{2}\check{\Delta}^{-1}\left[ d\widetilde{r}+\varepsilon
(\partial _{i_{1}}\ ^{1}\check{A})dx^{i_{1}}\right] ^{2}  \label{smallpol} \\
&&+(1+\varepsilon \chi _{7})\check{\Xi}\left[ d\widetilde{\vartheta }%
+\varepsilon (\partial _{i_{2}}\ _{1}^{2}n)dx^{i_{2}}\right]
^{2}+(1+\varepsilon \chi _{8})\mathring{h}_{5}\check{C}[d\psi +\varepsilon
(\partial _{i_{2}}\ ^{2}\check{A})dx^{i_{2}}+\ ^{1}\mathring{w}%
_{3}(x^{k})dt]^{2},  \notag
\end{eqnarray}%
where the $\varepsilon $--polarizations and N--coefficients are respectively
parameterized
\begin{eqnarray}
1+\varepsilon \chi _{3} &=&\check{\Phi}^{2}/4\Lambda \mathring{h}_{3}%
\mbox{
and }1+\varepsilon \chi _{4}=(\partial _{4}\check{\Phi})^{2}/\Lambda \check{%
\Phi}^{2}\mathring{h}_{4};  \label{aux6} \\
1+\varepsilon \chi _{5} &=&(\ ^{1}\check{\Phi})^{2}/4\ ^{1}\Lambda \check{A}%
\mbox{ and }1+\varepsilon \chi _{6}=(\partial _{6}\ ^{1}\check{\Phi})^{2}%
\check{\Delta}/\ ^{1}\Lambda (\ ^{1}\check{\Phi})^{2}\widetilde{r}^{2};
\notag \\
1+\varepsilon \chi _{7} &=&(\ ^{2}\check{\Phi})^{2}/4\ ^{2}\Lambda
\widetilde{r}^{2}\mbox{ and }1+\varepsilon \chi _{8}=(\partial _{8}\ ^{2}%
\check{\Phi})^{2}/\ ^{2}\Lambda (\ ^{2}\check{\Phi})^{2}\mathring{h}_{5}%
\check{C};\mbox{\ and }  \notag
\end{eqnarray}%
\begin{eqnarray*}
N_{i}^{3} &=&\check{N}_{i}^{3}=\check{n}_{i}(x^{k},\chi )=\varepsilon
\partial _{i}(\ _{1}n),\ N_{i}^{4}=\check{N}_{i}^{4}=\check{w}%
_{i}(x^{k},y^{4})=\partial _{i}\check{\Phi}/\partial _{4}\check{\Phi}; \\
N_{i_{1}}^{5} &=&\check{N}_{i_{1}}^{5}=\ ^{1}\check{n}_{i_{1}}(x^{k},\chi
,y^{6})=\varepsilon (\partial _{i_{1}}\ _{1}^{1}n),N_{i_{1}}^{6}=\check{N}%
_{i_{1}}^{6}=\varepsilon \ ^{1}\check{w}_{i_{1}}(x^{k},\chi ,y^{6})=\partial
_{i_{1}}\ ^{1}\check{\Phi}/\partial _{6}\ ^{1}\check{\Phi}; \\
N_{i_{2}}^{7} &=&\check{N}_{i_{2}}^{7}=\ ^{2}\check{n}_{i_{2}}(x^{k},\chi
,y^{6},y^{8})=\varepsilon (\partial _{i_{2}}\ _{1}^{2}n), \\
N_{i_{2}}^{8} &=&\delta _{i_{2}}^{3}\widetilde{N}_{3}^{8}+\delta _{i_{2}}^{5}%
\widetilde{N}_{5}^{8}+\check{N}_{i_{2}}^{8}=\delta _{i_{2}}^{3}\ ^{1}%
\mathring{w}_{3}+\varepsilon \ ^{2}\check{w}_{i_{2}}(x^{k},\chi
,y^{6})=\partial _{i_{2}}\ ^{2}\check{\Phi}/\partial _{8}\ ^{2}\check{\Phi}.
\end{eqnarray*}%
Such parametric solutions are also exact  for any fixed value of $%
\varepsilon .$ The generating and integration functions are not arbitrary
 but restricted to satisfy certain conditions of linear approximation on
$\varepsilon .$

Prescribing
\begin{equation}
\chi _{5}=2\overline{\zeta }\sin (\omega _{0}\chi +\chi _{0})
\label{ellipspol}
\end{equation}%
for constant parameters $\overline{\zeta },\omega _{0}$ and $\chi _{0},$ and
introducing the values $\ h_{5}=\check{A}(\widetilde{r})(1+\varepsilon \chi
_{5})=\widehat{A}(\widetilde{r},\chi )=2m(\chi )\widetilde{r}^{-2}-1,$ we
get an effective "anisotropically polarized" mass $m(\chi
)=m_{0}/[1+\varepsilon \overline{\zeta }\sin (\omega _{0}\chi +\chi _{0})].$
The condition $h_{5}=0$ results in a parametric formula for an ellipse with
eccentricity $\varepsilon ,$ $\widetilde{r}=2m_{0}/[1+\varepsilon \overline{%
\zeta }\sin (\omega _{0}\chi +\chi _{0})]$. Such ellipsoidal deformations of
Schwarzschild configurations in velocity space can be generated for a
special class of generating functions $\ ^{1}\check{\Phi}$ determined by
formulas (\ref{ellipspol}) and (\ref{aux6}). They result in effective
polarizations of the 4D components of the black ring metric modifying the
coefficients before $dt$ and $d\chi $ in (\ref{smallpol}).

\subsection{Time dependent solutions}

\noindent The classes of generic off--diagonal solutions constructed in the previous
sections are stationary  and do not depend on the time like coordinate $%
t,$ i.e. are with Killing symmetry on $\partial /\partial t.$ The AFDM
allows us to construct us exact solutions depending generically on time
variables. In a similar form, we can analogous solitonic configurations in
fiber spaces with evolution on $\tilde{t}.$ Solitonic solutions of gravitational field equations in (modified) gravity presents an important example of nonlinear wave interactions which may modify the background base spacetime and/or velocity/momentum type fibers. For deformations on a small parameters, such Finsler like solitonic metrics preserve the main physical properties of black ring type objects and describe certain additional vacuum polarizations and effective modifications of physical constants.

In this section, we analyze two
examples when velocity solitonic waves may result in observational effects
on the modified 4D Lorentz spacetime manifold.\\


\noindent (A.) We can choose  the generating function $\ ^{2}\check{\Phi}=\eta
(t,y^{6},y^{8})$ to be a solution of the Kadomtsev--Petviashvili (KP) equation
\cite{kp,vsol}
\begin{equation}
\pm \partial _{88}^{2}\eta +\partial _{6}(\partial _{t}\eta +\eta \partial
_{6}\eta +\epsilon \partial _{666}^{3}\eta )=0  \label{kdp}
\end{equation}%
where $\partial _{88}^{2}\eta :=\partial ^{2}\eta /\partial
y^{8}\partial y^{8}.$ and similarly for the other derivatives.
In a similar form, we can consider solutions of any 3--d solitonic
equations, for instance, generalized sine--Gordon ones. In the
dispersionless limit $\epsilon \rightarrow 0,$ we get solutions which do not
depend on $y^{8}$ but preserve locally anisotropic behaviour on $y^{6\text{\
}}$ and are determined by the Burgers' equation $\partial _{t}\eta +\eta \partial
_{6}\eta =0.$ Applying the same geometric method outlined in previous
section, we construct
\begin{eqnarray}
ds^{2} &=&e^{q}[\epsilon _{1}(dx^{1})^{2}+\epsilon _{2}(dx^{2})^{2}]+\frac{%
\check{\Phi}^{2}}{4\Lambda }[dt+\partial _{i}(\ _{1}n)dx^{i}]^{2}+\frac{%
(\partial _{4}\check{\Phi})^{2}}{\Lambda \check{\Phi}^{2}}[d\chi +(\partial
_{i}\check{A})dx^{i}]^{2}  \notag \\
&&+\frac{(\ ^{1}\check{\Phi})^{2}}{4\ ^{1}\Lambda }[d\widetilde{t}+(\partial
_{i_{1}}\ _{1}^{1}n)dx^{i_{1}}]^{2}+\frac{(\partial _{6}\ ^{1}\check{\Phi}%
)^{2}}{\ ^{1}\Lambda (\ ^{1}\check{\Phi})^{2}}[d\widetilde{r}+(\partial
_{i_{1}}\ ^{1}\check{A})dx^{i_{1}}]^{2}  \notag \\
&&+\frac{\eta ^{2}}{4\ ^{2}\Lambda }[d\widetilde{\vartheta }+(\partial
_{i_{2}}\ _{1}^{2}n)dx^{i_{2}}]^{2}+\frac{(\partial _{8}\ \eta )^{2}}{\
^{2}\Lambda \eta ^{2}}[d\psi +(\partial _{i_{2}}\ ^{2}\check{A}%
)dx^{i_{2}}]^{2},  \label{solit1}
\end{eqnarray}%
where there are certain differences comparing to (\ref{data1}) and (\ref%
{data2a}). For (\ref{solit1}), the polarization functions are
\begin{equation*}
\eta _{a_{2}}\mbox{  as }\eta _{7}=\eta ^{2}/4\ ^{2}\Lambda \widetilde{\Xi }%
\mbox{ and }\eta _{8}=\partial _{8}\eta ^{2}/\ ^{2}\Lambda \eta ^{2}%
\mathring{h}_{5}\widetilde{C},\ \forall \ \ \eta (t,y^{6},y^{8}),\partial
_{8}\ \forall \neq 0,
\end{equation*}%
and the N--coefficients are
\begin{eqnarray}
N_{i_{2}}^{7} &=&\check{N}_{i_{2}}^{7}=\ ^{2}\check{n}%
_{i_{2}}(t,y^{6},y^{8})=\ _{1}^{2}\check{n}_{i_{2}}(t,y^{6})+\ _{2}^{2}%
\check{n}_{i_{2}}(t,y^{6})\int d\psi (\partial _{8}\ \eta )^{2}/\eta ^{5},
\notag \\
N_{i_{2}}^{8} &=&\delta _{i_{2}}^{3}\widetilde{N}_{3}^{8}+\delta _{i_{2}}^{5}%
\widetilde{N}_{5}^{8}+\check{N}_{i_{2}}^{8}=\delta _{i_{2}}^{3}\ ^{1}%
\mathring{w}_{3}+\delta _{i_{2}}^{5}(\widetilde{B}^{2}/\widetilde{C})+\ ^{2}%
\check{w}_{i_{2}}(t,y^{6},y^{8})=\partial _{i_{2}}\ \eta /\partial _{8}\eta .
\notag
\end{eqnarray}%
Such solutions are not stationary, i.e. do not possess a Killing symmetry $%
\partial _{t}$ being different from those described by the quadratic element
(\ref{abrkerrbackgr1}). They define a self--consistent imbedding of a black
ring solution into velocity 3-d solitonic background determined by (\ref{kdp}%
). There is an off--diagonal interaction between the base and fiber degree
of freedoms via term $\ ^{1}\mathring{w}_{3}$ in $N_{i_{2}}^{8}.$

We can positively consider the metrics (\ref{solit1}) as solutions for
tangent Lorentz bundle black rings interacting with fiber solitons for some
small $\varepsilon $--deformations. For corresponding parameterizations, the
solution describe solitonic propagation of black rings in the total space.
In general, the nonlinear superposition of black ring and velocity solitonic
degrees of freedom do not have any obvious physical interpretation. \\

\vspace{.2 in}

\noindent (B.) The solitonic generating function $\ ^{2}\check{\Phi}=\check{\eta}(\tilde{t}%
,y^{6},y^{8})$ is considered to be a solution of the KP equation $\ \pm
\partial _{88}^{2}\check{\eta}+\partial _{6}(\partial _{\tilde{t}}\check{\eta%
}+\check{\eta}\partial _{6}\check{\eta}+\epsilon \partial _{666}^{3}\check{%
\eta})=0,$ with evolution on $\tilde{t}$ instead of $t$ which is different
from integral varieties in the previous example. This class of solitonic
solutions is defined by quadratic elements of type
\begin{eqnarray*}
ds^{2} &=&e^{\mathring{g}}[(dx^{1})^{2}-(dx^{2})^{2}]+(1+\varepsilon \chi
_{3})\mathring{h}_{3}\left[ dt+\varepsilon \partial _{i}(\ _{1}n)dx^{i}%
\right] ^{2}+(1+\varepsilon \chi _{4})\mathring{h}_{4}\left[ d\chi
+\varepsilon (\partial _{i}\check{A})dx^{i}\right] ^{2} \\
&&+(1+\varepsilon \chi _{5})\check{A}\left[ d\widetilde{t}+\varepsilon
(\partial _{i_{1}}\ _{1}^{1}n)dx^{i_{1}}\right] ^{2}+(1+\varepsilon \chi
_{6})\widetilde{r}^{2}\check{\Delta}^{-1}\left[ d\widetilde{r}+\varepsilon
(\partial _{i_{1}}\ ^{1}\check{A})dx^{i_{1}}\right] ^{2} \\
&&+(1+\varepsilon \chi _{7})\check{\Xi}\left[ d\widetilde{\vartheta }%
+\varepsilon (\partial _{i_{2}}\ _{1}^{2}n)dx^{i_{2}}\right]
^{2}+(1+\varepsilon \chi _{8})\mathring{h}_{5}\check{C}[d\psi +\varepsilon
(\partial _{i_{2}}\ ^{2}\check{A})dx^{i_{2}}+\ ^{1}\mathring{w}%
_{3}(x^{k})dt]^{2},
\end{eqnarray*}%
which are similar to (\ref{smallpol}) but with certain modifications of data
(\ref{aux6}) when
\begin{eqnarray*}
1+\varepsilon \chi _{7} &=&\check{\eta}^{2}/4\ ^{2}\Lambda \widetilde{r}^{2}%
\mbox{ and }1+\varepsilon \chi _{8}=(\partial _{8}\check{\eta})^{2}/\
^{2}\Lambda \check{\eta}^{2}\mathring{h}_{5}\check{C};\mbox{\ and } \\
N_{i_{2}}^{8} &=&\delta _{i_{2}}^{3}\widetilde{N}_{3}^{8}+\delta _{i_{2}}^{5}%
\widetilde{N}_{5}^{8}+\check{N}_{i_{2}}^{8}=\delta _{i_{2}}^{3}\ ^{1}%
\mathring{w}_{3}+\varepsilon \ ^{2}\check{w}_{i_{2}}(\tilde{t}%
,y^{6},y^{8})=\partial _{i_{2}}\ \check{\eta}/\partial _{8}\ \check{\eta}.
\end{eqnarray*}%
Such small solitonic deformations in velocity variables preserve
stationarity on 4D base Lorentz manifold. Nevertheless,
interactions between the base and fibers is given by the term $\ ^{1}%
\mathring{w}_{3}$ in $N_{i_{2}}^{8}.$ Such black ring and/or (modified) gravitational solitonic objects  are not subject to restrictions of
Michelson--Morley type experiments for locally anisotropic aether \cite{lam}
being constructed as potential relativistic astrophysical objects in compact
spacetime regions.
 Additional
constraints are needed to be imposed in order to select LC--configurations.

\section{Concluding Remarks}

\label{s5} Black ring objects studied in this paper can be real and can be the result of
 QG effects via nonlinear polarizations in 4D gravity theories even if
the space time lacks extra dimensions in the sense of Kaluza-Klein.

Locally anisobropic black rings
may exist as 4D osculatory "shadows" of MGTs with arbitrary MDRs. On
tangent Lorentz bundles there are no restrictions on  black hole uniqueness
theorems. This paves the way to construct various classes of black ellipsoid/hole,
cosmological and solitonic solutions of Finsler modified gravitational field
equations \cite{vfinslbranes,vfinlbh,stavrv1}. Such objects may have
 cosmological implications, for instance, if galatic nucleus may
contain analogous black torus solutions with dark matter and dark energy. The AFDM developed in this work can also be applied  to construct similar generic
off--diagonals solutions with extra dimensions in  brane and string gravity.

Physical interpretation of Finsler black ring and black hole solutions depends on the type of nonholonomic constraints we impose, i.e. on the class of generating and integration functions and constants we chose to define our solutions and on resulting off-diagonal deformations and effective polarizations of constants.  For small parametric N-connection coefficients and small deformations of standard black ring/hole solutions, we positively get high dimensional black hole solutions with extra "velocity" type coordinates. In such cases, there are typical horizons and singularities similar to the case of black holes in higher dimension theories. It is possible to construct toy $2+2$ dimensional Finsler like black hole/ellipsoid models with two "velocity" coordinates with smooth nonholnomic imbedding in a five dimensional spacetime. For such configurations, we can speculate on analogous censorship theorems but these speculations apply only to the specific models constructed.  It is also possible to generate solutions for naked singularities imbedded into solitonic like backgrounds \cite{vsol}, with commutative and noncommutative Finsler brane warping/trapping on velocity type coordinates \cite{vgrg,vfinslbranes,vfinlbh} and nontrivial nonlinear connection structure.

Let us discuss the Lorentz invariance violation that is considered to be a general property of Finsler like theories. This is not always correct to assume Lorentz invariance violation is a general property of Finsler like theories because  Einstein’s theory of gravity can be described also by Finsler like variables by prescribing a nonholonomic 2+2 splitting on Lorentz manifolds when the local Lorentz invariance is implicit in such constructions\cite{veinstf1,veinstf2,veinstf3}.\footnote{A pseudo-Riemannian manifold can be described in terms of any connection which is defined by a distortion tensor from the standard Levi-Civita connection, for instance, using the Cartan Finsler connection for a fibered structure. Sure, such constructions involve nontrivial torsion and non-metricity geometric objects but the corresponding distortion tensor is determined by a Sasaki type metric structure and conventional N--connection coefficients induced by generic off-diagonal metric terms. These are the so-called Finlser like, or almost Kaehler-Finsler like variables which can be used for the deformation quantization of the Einstein gravity and generating various classes of black hole and cosmological solutions. We can always re-encode all constructions in terms of standard metric and Levi-Civita variables, or introduce generalized tetradic, diadic, spinor variables etc.}.  Nevertheless, we always get Lorentz violations if we work with theories on tangent bundles. For compatibility with standard particle physics and gravity theories, we need to consider tangent Lorentz bundles. We can preserve for certain configurations a conventional local Lorentz invariance, which is modified by certain contributions from gravitational interactions depending on spacetime and velocity/momentum type variables. Here we note that even in general relativity, Lorentz transforms are considered only for fixed spacetime points but Lorentz invariance does not apply generally. A similar property exists for the Einstein-Finsler type theories on tangent Lorentz bundles when a corresponding axiomatic can be formulated in a form similar to that for the general relativity theory,  see \cite{vaxiom}. Violation of the Lorentz symmetry in such Cartan-Finsler type generalized models is a consequence of locally anisotropic gravitational and matter field interactions depending both on base Lorentz manifold coordinates and on fiber like (velocity/momentum) coordinates on (co) tangent bundles.

 Various schools on Finsler geometry and generalizations usually work with more general modifications of the Einstein gravity (in many cases, such models do not limit   to Einstein’s gravity, e.g., see critical remarks and references in  \cite{vcritics}). The first class of such generalized models involve certain modified local Lorentz transforms (broken local Lorentz invariance).  Finsler metrics can be related to modified dispersion relations like in Refs. \cite{lam,vgrg} and, for the second class of generalizations, various types of Finsler geometric models can be elaborated upon for different classes of Finsler connections (which may or may not be metric compatible).  Such "physical" theories are constructed on different principles than the standard theories of gravity. We could formulate self-consistent commutative and non-commutative  Finsler generalizations of Einstein gravity using equivalent geometric variables with the so-called canonical d-connection and the Cartan-Finsler connection. In such cases, it is possible to solve generalized Einstein-Finsler equations in exact forms and derive various classes of generalized black ring/ ellipsoid/  hole and wormhole solutions as in the present work and in \cite{vgrg,vfinslbranes,vfinlbh} by applying geometric methods summarized in Ref. \cite{vkerrhd}.

Finally, we discuss a recent work on exact solutions in a model of Finsler spacetime  \cite{finsexsol}, which is elaborated upon as an alternative to the models with the Cartan and canonical d-connection. The authors work directly with a Ricci like  tensor in Finsler geometry introduced by Akbar-Zadeh \cite{akbar} which can be defined in symmetric form just from the fundamental Finsler function $F$ not involving the concept of linear connection. Such an approach is based on nice geometric constructions. Nevertheless, it does not provide a self-consistent physical theory without additional motivations for a covariant derivative (linear connection), which should be adapted to the nonholonomic structure. There are also necessary some postulates on nonholonomic frames and physical observers, Lagrangians for anisotropic gravitational and matter field interactions. In our opinion \cite{vaxiom,vcritics} such theories should be with curvature and Ricci tensors for the Cartan--Finsler and/or canonical d-connection. Akbar-Zadeh configurations can be extracted at the end via corresponding classes of nonholonomic deformations of fundamental geometric objects. For anisotropic interactions with matter fields, this is a very difficult theoretical problem.

In general relativity, such issues are solved in a "simplified" way due to the existence of the Levi-Civita connection and the formulation of  the general principle of relativity (for arbitrary frames of reference). A physically viable Finsler spacetime geometry can not be derived only from the generalized Finsler metric $F$, or its Hessian, or  semi-spray function etc. We need additional assumptions on linear and nonlinear connections (with frame adapted structures) which can be motivated following certain geometric physical principles. In a more general context, it is necessary to formulate some self-consistent generalized gravitational and matter field equations for a model of Finsler gravity which would be compatible with Einstein’s gravity. There are necessary some explicit exact solutions and quantum models for Finsler like gravity theories. In a more advanced theoretical framework such constructions were elaborated upon by using the almost symplectic Cartan-Finsler connection and the canonical d-connection, see details in Refs. \cite{vcritics,vaxiom}. Following such an approach, it was possible to construct physically important black hole like and cosmological exact solutions \cite{vkerrhd,vgrg,vfinslbranes,stavrv1}  with various supersymmetric, superstring and nocommutative generalizations \cite{vacarunp1,vfinlbh}. To conclude, Finsler type theories can be important in modern cosmology and astrophysics, and various quantum gravity models, because of global and local anisotropic classical and possible quantum effects.  However, classical Lorentz violating terms must be small in order to have compatibility with the data for our Solar System. In our work such an assumption is implicit.
\vskip5pt

\textbf{Acknowledgments:\ } SV work is partially supported by IDEI,
PN-II-ID-PCE-2011-3-0256, and a DAAD fellowship.


\begin{thebibliography}{99}

\bibitem{1} H. Elvang, R. Emparan, D. Mateos
and H. S. Reall, A supersymmetric black ring, Phys. Rev. Lett. 93 (2004) 211302;\ arXiv: hep-th/0407065

\bibitem{2} H. Elvang, R. Emparan, D. Mateos, and H. S. Reall, Supersymmetric black rings and
three-charge supertubes, Phys.Rev. D71 (2005) 024033;\ arXiv: hep-th/0408120

\bibitem{emp} R. Emparan and H. S. Reall, A rotating black ring in five
dimensions, Phys. Rev. Lett. 88 (2002) 101101;\ arXiv: hep-th/0110260

\bibitem{emp1} R. Emparan and H. S. Reall, Black rings, Class. Quant. Grav.
23 (2006) R169;\ arXiv: hep-th/0608012

\bibitem{15} A. A. Pomeransky and R. A. Sen’kov, Black ring with two angular momenta, arXiv: hep-th/0612005

\bibitem{16} H. Elvang, P. Figueras, Black Saturn, JHEP 0705, 050 (2007);\ arXiv: hep-th/0701035

\bibitem{17} H. Iguchi and T. Mishima, Black di-ring and infinite nonuniqueness, Phys. Rev. D 75, 064018 (2007)
[Erratum-ibid. D 78, 069903 (2008)];\ arXiv: hep-th/0701043

\bibitem{18} J. Evslin and C. Krishnan, The Black Di-Ring: An Inverse Scattering Construction, Class. Quant.
Grav. 26, 125018 (2009);\ arXiv:0706.1231

\bibitem{19} H. Elvang and M. J. Rodriguez, Bicycling Black Rings, JHEP 0804, 045 (2008);\ arXiv:0712.2425

\bibitem{20} Y. Chen, K. Hong and E. Teo, “Unbalanced Pomeransky-Sen’kov black ring,” Phys. Rev. D 84, 084030
(2011);\ arXiv:1108.1849

\bibitem{21} J.V. Rocha and M.J. Rodriguez, unpublished.

\bibitem{22} Y. Chen and E. Teo, Black holes on gravitational instantons, Nucl. Phys. B 850, 253 (2011);\
arXiv:1011.6464 [hep-th]

\bibitem{23} Y. Chen and E. Teo, Rotating black rings on Taub-NUT, arXiv: 1204.3116 [hep-th]
\bibitem{A} H. Elvang, R. Emparan, D. Ma-
teos and H. S. Reall, Supersymmetric black rings and three-charge supertubes, Phys.
Rev. D71 (2005) 024033; [hep-th/0408120]. H. Elvang, R. Emparan, D. Mateos
and H. S. Reall, A supersymmetric black ring, Phys. Rev. Lett. 93 (2004) 211302;
[hep-th/0407065].
\bibitem{B} J. B. Gutowski and H. S. Reall, Supersymmetric AdS5 Black Holes, JHEP 02 (2004)
006; [hep-th/0401042].
\bibitem{BB} R. Emparan and H. S. Reall, “Black rings,” Class. Quant. Grav. 23, R169 (2006)
[arXiv:hep-th/0608012].
\bibitem{C} J. B. Gutowski and H. S. Reall, General Supersymmetric AdS5 Black Holes, JHEP
04 (2004) 048; [hep-th/0401129].
\bibitem{D} D. Klemm and W. A. Sabra, Supersymmetry of Black Strings in D = 5 Gauged
Supergravities, Phys. Rev. D62 (2000) 024003; [hep-th/0001131].
\bibitem{E} J. P. Gauntlett and J. B. Gutowski, All Supersymmetric Solutions of Minimal Gauged
Supergravity in Five Dimensions, Phys. Rev. D68 (2003) 105009; [hep-th/0304064].
\bibitem{F} J. B. Gutowski and W. A. Sabra, General Supersymmetric Solutions of Five-
Dimensional Supergravity, JHEP 10 (2005) 039; [hep-th/0505185].
\bibitem{G} A. Bouchareb, G. Clement, C. M. Chen, D. V. Gal’tsov, N. G. Scherbluk and T. Wolf, “G2
generating technique for minimal D=5 supergravity and black rings,” Phys. Rev. D 76,
104032 (2007) [arXiv:0708.2361 [hep-th]].
\bibitem{vaxiom} S. Vacaru, Principles of Einstein-Finsler gravity and
perspectives in modern cosmology, Int. J. Mod. Phys. D 21 (2012) 1250072 (40
pages); arXiv: 1004.3007

\bibitem{stavrv1} P. Stavrinos and S. Vacaru, Cyclic and ekpyrotic universes
in modified Finsler osculating gravity on tangent Lorentz bundles, Class.
Quant. Grav. 30 (2013) 055012; arXiv: 1206.3998

\bibitem{vkerrhd} T. Gheorghiu, O. Vacaru and S. Vacaru, Off--diagonal
deformations of Kerr black holes in Einstein and modified massive gravity
and higher dimensions,  EPJC 74 (2014) 3152;\ arXiv: 1312.4844

\bibitem{odintsov} S. Nojiri and S. D. Odintsov, Unfied cosmic history in
modified gravity: From F(R) theory to Lorentz non--ivariant models, Phys.
Rept. 505 (2011) 59-144;\ arXiv: 1011.0544

\bibitem{capozzello} S. Capozzello and V. Fraoni, Beyond Einstein Gravity: A
Survey of Gravitational Theories for Cosmology and Astrophysics, Fundamental
Theories of Physics, vol. 170, (Springer Netherelands, 2011), 467 pp

\bibitem{vgrg} S. Vacaru, Modified dispersion relations in Horava-Lifshitz
gravity and Finsler brane models, Gener. Relat. Grav. 44 (2012) 1015-1042;\
arXiv: 1010.5457

\bibitem{vfinslbranes} S. Vacaru, Finsler branes and quantum gravity
phenomenology with Lorentz symmetry violations, Class. Quant. Grav. 28
(2011) 215991; arXiv: 1008.4912

\bibitem{vfinlbh} S. Vacaru, Finsler black holes induced by noncommutative
anholonomic distributions in Einstein gravity, Class. Quant. Grav. 27 (2010)
105003; arXiv: 0907.4278

\bibitem{vacarunp1} S. Vacaru, Superstrings in higher order extensions of
Finsler superspaces, Nucl. Phys. B, 434 (1997) 590-656;\ arXiv:
hep-th/9611034


\bibitem{twotime1} I. Bars, Shih-Hung Chen, Geometry and symmetry structures
in 2T gravity, Phys. Rev. D 79 (2009) 085021, arXiv: 0811.2510

\bibitem{twotime2} C. Castro, Gravity in curved phase-spaces, Finsler
geometry and two--times physics, Int. J. Mod. Phys. A. 27 (2012) 418-430

\bibitem{twotime3} J. A. Nieto, Towards a canonical gravity in two time and
two space dimensions, Int. J. Geom. Meth. Mod. Phys. 09 (2012) 1250069

\bibitem{yano} K. Yano and S. Ishihara, Tangent and Contangent Bundles (M.
Dekker, 1973)

\bibitem{heusler} M. Heusler, Black Hole Uniqueness Theorems, (Cambridge
University Press, 1996)

\bibitem{kramer} D. Kramer, H. Stephani, E. Herdlt and M. A. H. MacCallum,
Exact Solutions of Einstein's Field Equations, 2d edition (Cambridge
University Press, 2003)

\bibitem{misner} C. W. Misner, K. S. Thorne and J. A. Wheeler, Gravitation
(Freeman, 1973)

\bibitem{kp} B. B. Kadomtsev and V. I. Petviashvili, On the stability of
solitary wavew in weekly dispersive media Doklady Akademii Nauk SSS 192 (1970) 753-756 [Russian]; Sov. Phys. Dokl. 15 (1970) 539-541 [English]

\bibitem{vsol} S. Vacaru, Curve flows and solitonic hierarchies generated by
Einstein metrics, Acta Applicandae Mathematicae 110 (2010) 73-107

\bibitem{lam} C. Lammerzahl, D. Lorek and H. Dittus, Confronting Finsler
spacetime with experiment, Gen. Rel. Grav.  41  (2009) 1345-1353; arXiv: 0811.0282

\bibitem{veinstf1} S. Vacaru,  Parametric nonholonomic frame transforms and exact solutions in gravity, Int. J. Geom. Meth. Mod. Phys.  4  (2007) 1285-1334;  arXiv: 0704.3986 [gr-qc]

\bibitem{veinstf2} S. Vacaru,   Finsler and Lagrange geometries in Einstein and string gravity,  Int. J. Geom. Meth. Mod. Phys.  5  (2008) 473-511; arXiv: 0801.4958 [gr-qc]

 \bibitem{veinstf3}   S. Vacaru,  Einstein gravity as a nonholonomic almost Kaehler geometry, Lagrange-Finsler variables, and  deformation quantization, J. Geom. Phys.  60  (2010) 1289-1305; arXiv: 0709.3609 [math-ph]

\bibitem{vcritics} S. Vacaru, Critical remarks on Finsler modifications of gravity and cosmology by Zhe Chang and Xin Li, Phys. Lett. B 690 (2010) 224-228; arXiv: 1003.0044v2 [gr-qc]

\bibitem{finsexsol} Xin Li and Zhe Chang, An exact soluton of vacuum field equation in Finsler spacetime, Phys. Rev. D 90 (2014) 064049;  arXiv: 1401.6363

\bibitem{akbar} H. Akbar-Zadeh, Sur les espaces de Finsler a courbures sectionnelles constantes, Acad. Roy. Belg, Bull. CI. Sci (5) 74 (1988) 281-322
\end{thebibliography}
\end{document}